\documentclass[prb,twocolumn,showpacs]{revtex4}
\usepackage{graphicx}
\usepackage{amsfonts,amsmath,amssymb,float}
\usepackage{bm,dsfont}
\usepackage{color}
\usepackage{bbm}
\usepackage{longtable}
\usepackage[dvips]{epsfig}
\usepackage{amsmath,amssymb,lscape,float}
\usepackage{hyperref}
\usepackage{listings}

\def\tr{\qopname\relax{no}{Tr}}

\begin{document}
\title{Entanglement-spectrum characterization of ground-state nonanalyticities in 
coupled excitation-phonon models}

\author{Vladimir M. Stojanovi\'c}
\email{vladimir.stojanovic@physik.tu-darmstadt.de} 
\affiliation{Institut f\"{u}r Angewandte Physik, Technical
University of Darmstadt, D-64289 Darmstadt, Germany}

\date{\today}

\begin{abstract} 
The polaron concept captures physical situations involving an itinerant quantum particle (excitation) that interacts strongly 
with bosonic degrees of freedom and becomes heavily boson-dressed. While the Gerlach-L\"{o}wen theorem rules out the occurrence of 
nonanalyticities of ground-state-related quantities for a broad class of polaron models, examples were found in recent years of 
sharp transitions pertaining to strongly momentum-dependent interactions of an excitation with dispersionless (zero-dimensional) 
phonons. On the example of a lattice model with Peierls-type excitation-phonon interaction, such level-crossing-type small-polaron 
transitions are analyzed here through the prism of the entanglement spectrum of the excitation-phonon system. By evaluating this 
spectrum in a numerically-exact fashion it is demonstrated that the behavior of the entanglement entropy in the vicinity of the 
critical excitation-phonon coupling strength chiefly originates from one specific entanglement-spectrum eigenvalue, namely the 
smallest one. While the discrete translational symmetry of the system implies that those eigenvalues can be labeled by the bare-excitation 
quasimomentum quantum numbers, here it is shown numerically that they are predominantly associated to the quasimomenta $0$ and $\pi$, 
including cases where a transition between the two takes place deeply in the strong-coupling regime.
\end{abstract}

\maketitle
\section{Introduction}
Recent years have witnessed an ever-increasing proliferation of techniques from 
quantum-information theory into the field of condensed-matter physics~\cite{Laflorencie:16}. 
The first major surge of interest in this direction entailed the use of bipartite entanglement and 
the concept of entanglement entropy to characterize various quantum phase transitions~\cite{Amico:08}, 
primarily in the realm of strongly-correlated and quantum-spin systems. For a quantum system that can be 
partitioned into two entangled subsystems, the entanglement entropy is defined as the von Neumann entropy 
of the reduced density matrix pertaining to either one of the two subsystems, obtained by tracing out the 
degrees of freedom of the other one. This entropy -- a single number -- represents a quantitative measure 
of entanglement in any given state of a bipartite quantum system.

Over the past decade, the concept of the {\em entanglement spectrum} attracted considerable interest in the 
context of the symmetry-protected topological states of matter~\cite{Li+Haldane:08}. It arises naturally -- simply 
by noticing that each reduced density matrix can be written in the form $\rho=\exp(-H_{\textrm{E}})$, i.e., as the 
canonical density matrix corresponding to a ``Hamiltonian'' $H_{\textrm{E}}$ at the inverse temperature 
$\beta_{\textrm{E}}=1$~\cite{Chandran+:14}. In the same vein, the entanglement entropy can be thought of as the 
thermodynamic entropy~\cite{Wehrl:78,HayashiBOOK} of a system described by $H_{\textrm{E}}$. This last 
Hamiltonian, the negative logarithm of the reduced density matrix, became known as the modular (or entanglement) 
Hamiltonian and the set of its eigenvalues the entanglement spectrum. Such spectra have 
already proven their worth as they were shown to capture the edge physics of topologically ordered 
phases~\cite{Li+Haldane:08,Pollmann+:10,Thomale+:10}, a research direction pioneered by Li and Haldane~\cite{Li+Haldane:08}. 
They also led to nontrivial physical insights in other condensed-matter areas, e.g., interacting spin chains~\cite{Poilblanc:10,Laeuchli+Schliemann:12}, 
topological insulators and superconductors~\cite{Fidkowski:10}, integer quantum Hall effect~\cite{Schliemann:11}, interacting 
bosons~\cite{Deng+Santos:11,Ejima+:14} and fermions~\cite{Toldin+Assad:18}, the Hofstadter problem~\cite{Huang+Arovas:12,Schliemann:13}, 
and many-body localization~\cite{Yang+:15,Geraedts+:17}. This can be attributed to the fact that entanglement spectrum 
provides a more detailed characterization of the pattern of entanglement in a given system than the corresponding entropy~\cite{Li+Haldane:08}.

One area of condensed-matter physics that has not been explored yet from the entanglement-spectrum viewpoint is that of small 
polarons~\cite{Emin:82,RanningerReview:06,AlexandrovDevreeseBook} -- quasiparticles emerging in lattice models based on the 
molecular-crystal paradigm~\cite{Holstein:59}. Those models describe a short-ranged coupling of an itinerant excitation to 
dispersionless phonons~\cite{Engelsberg+Schrieffer:63}, representing an abstraction for the physical situation in which an excess 
charge carrier or an exciton in a crystal of a narrow-band semiconductor (or an insulator) interacts with optical phonons of 
the host crystal. A strong excitation-phonon (e-ph) coupling leads to a heavily phonon-dressed excitation, that acquires at the same 
time a large effective band mass. In particular, if the spatial extent of its wave function does not exceed one unit cell of 
the host crystal the ensuing phonon-dressed quasiparticle is referred to as small polaron~\cite{Emin:82}.

Naturally arising from investigations of transport properties of narrow-band electronic 
materials~\cite{Hannewald++:04,Slezak++:06,Fratini+Ciuchi:09,StojanovicGraphene}, in the 
course of time studies of small-polaron models spawned a research area important in its own 
right~\cite{Ranninger:92,Capone+:97,Wellein+Fehske,Jeckelmann+White:98,Bonca+:99,Zoli:03,Stojanovic+:04,
Edwards:06,Alvermann:07,Stojanovic+:12,Mei+:13,Chakraborty+:16,Jansen+:19}. 
While the bulk of such studies have been devoted to the time-honored Holstein model~\cite{Holstein:59}, 
which captures the dependence of the excitation's on-site energy upon Einstein-phonon displacements 
on the same site (local e-ph coupling), over the past two decades considerable attention was devoted to various models 
with nonlocal-coupling mechanisms~\cite{Zoli:03,Stojanovic+:04,Edwards:06,Alvermann:07}. The most well known among them 
is the Peierls-coupling mechanism (also known as Su-Schrieffer-Heeger- or off-diagonal coupling)~\cite{Stojanovic:08}, 
which accounts for the effective dependence of the hopping amplitude between adjacent lattice sites upon the difference of 
local Einstein-phonon displacements on those sites. 

An important point of distinction between various coupled e-ph models is provided by the Gerlach-L\"{o}wen 
theorem~\cite{Gerlach+Lowen:87,GerlachLowenRMP:91}. This rigorous result rules out nonanalyticities in ground-state-related properties 
for all models with e-ph vertex functions that are either completely momentum-independent (Holstein-type coupling~\cite{Holstein:59}) 
or depends on the phonon quasimomentum, but not on that of the excitation (e.g., Fr\"{o}hlich-type coupling~\cite{Froehlich:54}). 
Thus, couplings that depend on both the excitation and phonon quasimomenta do not belong to the domain of applicability of 
this theorem. Moreover, for some particular e-ph interactions of this type -- with the Peierls-type coupling being the prime 
example -- level-crossing-type sharp transitions were shown to exist~\cite{Stojanovic:08,Sous+:17,Stojanovic+:14}. Namely, 
at certain critical coupling strengths their ground states change their character from nondegenerate zero-quasimomentum ones to 
twofold-degenerate ones corresponding to a symmetric pair of nonzero quasimomenta. To demonstrate one such transition, a 
quantum simulator based on superconducting qubits and resonators was proposed~\cite{Stojanovic+:14,Stojanovic+Salom:19}.

In this paper, the sharp transition in a one-dimensional (1D) model with Peierls-type coupling is analyzed from 
the point of view of the entanglement spectrum of the underlying (bipartite) e-ph system. In particular, the main aim of
this paper is to analyze the dependence of the entanglement-spectrum eigenvalues on the effective e-ph coupling strength. 
Its principal finding is that the behavior of the entanglement entropy in the vicinity of the critical coupling strength is 
to a large extent determined by the smallest eigenvalue. It is also demonstrated that -- as a consequence of the discrete translational 
symmetry of the system -- the eigenvalues from the entanglement spectrum can be labeled by the bare-excitation quasimomentum 
quantum numbers. This is complemented by the numerical finding that this quantum number in the model under consideration takes 
values $0$ and $\pi$, including cases where a transition between the two occurs at a coupling strength far larger 
than the critical one.

The remainder of this paper is organized as follows. In Sec.~\ref{ModelMethod} the relevant coupled e-ph Hamiltonian is introduced
(Sec.~\ref{ModelHamiltonian}), along with a short description of the computational methodology utilized here to compute its ground-state 
properties (Sec.~\ref{Methodology}). In Sec.~\ref{entspectrum}, after recapitulating the most general properties of entanglement in 
bipartite systems (Sec.~\ref{Bipartite}), basic aspects of entanglement spectra in such systems are briefly reviewed (Sec.~\ref{EntSpectrum}), 
followed by some general symmetry-related considerations and their specific application to the coupled e-ph system at hand 
(Sec.~\ref{symmconsid}). The main findings of this paper are presented and discussed in Sec.~\ref{ResultsDiscuss}. Finally, the paper 
is summarized, with conclusions and some general remarks, in Sec.~\ref{SumConcl}. An involved mathematical derivation is relegated to 
Appendix~\ref{KeExpr}.
\section{Model and method} \label{ModelMethod}
\subsection{Hamiltonian of the system} \label{ModelHamiltonian}
The system under consideration comprises a spinless-fermion excitation nonlocally coupled to dispersionless 
phonons. It is described by a 1D e-ph model, whose Hamiltonian can succinctly be written as
\begin{equation} \label{Hamiltonian}
H = H_{\rm e} + H_{\rm ph} + H_{\textrm{e-ph}}  \:.
\end{equation}
Here $H_{\rm e}$ is the excitation hopping (i.e., kinetic-energy) term in the tight-binding 
representation, given by
\begin{equation}
H_{\rm e} = -t_{\rm e}\sum_n (c^\dagger_{n+1}c_n + \mathrm{H.c.}) \:,
\end{equation}
with $t_{\rm e}$ being the corresponding hopping amplitude; $c_{n}^{\dagger}$ ($c_{n}$) creates (destroys) an 
excitation at site $n$ ($n=1,\ldots,N$). [For simplicity, the excitation on-site energy is set to zero in the 
following.] At the same time $H_{\rm ph}$ stands for the free-phonon term ($\hbar=1$ in what follows)
\begin{equation}
H_{\rm ph} = \omega_{\textrm{ph}} \sum_n b^\dagger_n b_n \:,
\end{equation}
where $b_{n}^{\dagger}$ ($b_{n}$) creates (destroys) an Einstein phonon with frequency $\omega_{\textrm{ph}}$ at site $n$. 
Finally, the e-ph coupling term describes the lowest-order (linear) dependence of the effective hopping amplitude 
between two adjacent sites, say $n$ and $n+1$, on the difference of the local phonon displacements $u_{n+1}$ and 
$u_n$ (where $u_n\propto b^\dagger_n + b_n$) at those sites (Peierls-type coupling). It is given by
\begin{equation}\label{Heph}
H_{\textrm{e-ph}} = g\omega_{\textrm{ph}} \sum_n (c^\dagger_{n+1}c_n + \mathrm{H.c.})
(b^\dagger_{n+1} + b_{n+1} - b^\dagger_n - b_n) \:,
\end{equation}
with $g$ being the dimensionless coupling strength.

The eigenstates of the Hamiltonian $H$ in Eq.~\eqref{Hamiltonian} ought to be good-quasimomentum states, i.e., 
joint eigenstates of $H$ and the total quasimomentum operator
\begin{equation}\label{totalcryst}
K_{\textrm{tot}}=\sum_{k} k\:c^{\dagger}_{k}c_{k}+\sum_{q}q\:b^{\dagger}_{q}b_{q} \:,
\end{equation}
since the latter commutes with $H$. In the following, the eigenvalues of $K_{\textrm{tot}}$ are labelled with $K$ and 
quasimomenta are dimensionless, i.e., expressed in units of the inverse lattice period. In particular, use is made of 
periodic boundary conditions, with $N$ permissible quasimomenta in the Brillouin zone given by $k_n=2\pi n/N$, 
where $n=-N/2+1,\ldots, N/2$ ($N$ is assumed to be even).

By switching to momentum space, the e-ph coupling Hamiltonian of Eq.~\eqref{Heph} can be recast in the generic form
\begin{equation}\label{mscoupling}
H_{\mathrm{e-ph}}=N^{-1/2}\sum_{k,q}\gamma_{\textrm{e-ph}}(k,q)\:
c_{k+q}^{\dagger}c_{k}(b_{-q}^{\dagger}+b_{q}) \:,
\end{equation}
where its corresponding vertex function is given by
\begin{equation}\label{vertex_func}
\gamma_{\textrm{e-ph}}(k,q)=2ig\:\omega_{\textrm{ph}}\:[\:\sin k-\sin(k+q)] \:.
\end{equation}
Because the latter depends both on $k$ and $q$, the Peierls-coupling term in Eq.~\eqref{Heph} does not satisfy 
the conditions for the applicability of the Gerlach-L\"{o}wen theorem~\cite{GerlachLowenRMP:91}.

Ground-state properties of small polarons are customarily discussed in terms of an effective coupling strength. 
For the most general (momentum-dependent) vertex function $\gamma_{\textrm{e-ph}}(k,q)$, the effective coupling 
strength is defined as $\lambda_{\textrm{eff}}=\langle|\gamma_{\textrm{e-ph}}(k,q)|^{2}\rangle_{\textrm{BZ}}/(2t_{\rm e}\:
\omega_{\textrm{ph}})$, where $\langle\ldots\rangle_{\textrm{BZ}}$ stands for the Brillouin-zone average. For 
$\gamma_{\textrm{e-ph}}(k,q)$ given by Eq.~\eqref{vertex_func}, this reduces to $\lambda_{\textrm{eff}}\equiv 
2g^{2}\:\omega_{\textrm{ph}}/t_{\rm e}$. In particular, the ground state of the Hamiltonian \eqref{Hamiltonian} 
with Peierls-type coupling undergoes a sharp level-crossing-type transition (i.e., first-order nonanalyticity) at 
a critical value $\lambda^{\textrm{c}}_{\textrm{eff}}\sim 1$ of $\lambda_{\textrm{eff}}$~\cite{Stojanovic:08,Sous+:17}. 
For $\lambda_{\textrm{eff}}<\lambda^{\textrm{c}}_{\textrm{eff}}$ the ground state is the (nondegenerate) $K=0$ eigenvalue 
of $K_{\mathrm{tot}}$, while for $\lambda_{\textrm{eff}}\ge\lambda^{\textrm{c}}_{\textrm{eff}}$ it is twofold-degenerate 
and corresponds to a symmetric pair of nonzero quasimomenta $K=\pm K_{\textrm{gs}}$. Upon increasing $\lambda_{\textrm{eff}}$ 
beyond its critical value, $K_{\textrm{gs}}$ also changes -- which is reflected in the ground-state energy as a sequence 
of further first-order nonanalyticities -- and saturates at $K_{\textrm{gs}}=\pi/2$ for a sufficiently large $\lambda_{\textrm{eff}}$. 
Importantly, both $\lambda^{\textrm{c}}_{\textrm{eff}}$ and the values of $\lambda_{\textrm{eff}}$ that correspond 
to the latter nonanalyticities are not universal, being dependent on the adiabaticity ratio $\omega_{\textrm{ph}}/t_{\rm e}$.

It is worthwhile to mention that a similar sharp transition was found~\cite{Stojanovic+:14,Stojanovic+Salom:19} in a model where 
Peierls-type coupling is complemented by e-ph interaction of the breathing-mode type~\cite{Slezak++:06}. It is important to  
stress that a dependence on both the excitation and phonon quasimomenta $(k,q)$ is not a sufficient condition for the existence 
of a ground-state nonanalyticity; a counterexample is furnished, e.g., by the Edwards model~\cite{Edwards:06,Alvermann:07,Chakraborty+:16}. 
\subsection{Computational methodology} \label{Methodology}
The ground-state properties of the e-ph system at hand are here computed using the conventional Lanczos diagonalization 
method for sparse matrices~\cite{CullumWilloughbyBook,PrelovsekBoncaChapter:13}, combined with a controlled truncation of 
the (otherwise infinite-dimensional) phonon Hilbert space. 

The Hilbert space of the e-ph system is spanned by states of the form $|n\rangle_{\textrm{e}} \otimes |\mathbf{m}\rangle_\text{ph}$,
where $|n\rangle_{\textrm{e}}\equiv c_{n}^{\dagger}|0\rangle_{\textrm{e}}$ represents an excitation localized at site 
$n$, $\mathbf{m}\equiv(m_1,\ldots,m_N)$ is the set of phonon occupation numbers, and $|\mathbf{m}\rangle_\text{ph} =
\prod_{i=1}^N(1/\sqrt{m_i!})(b_i^\dagger)^{m_i}|0\rangle_\text{ph}$ (here $|0\rangle_{\textrm{e}}$ and $|0\rangle_{\textrm{ph}}$
are the excitation and phonon vacuum states, respectively). With the restriction to a truncated phonon space comprising
states with at most $M$ phonons, all $m$-phonon states with $0\le m_i \le m$ are included, where $m=\sum_{i=1}^N m_i \le M$. 
The dimension of the total Hilbert space is given by $D = D_\text{e} \times D_\text{ph}$, where $D_\text{e} = N$ and $D_\text{ph}=
(M+N)!/(M!N!)$. A generic state in this Hilbert space is given by
\begin{equation}\label{expandPsi}
|\psi\rangle=\sum_{n,\mathbf{m}}C_{n,\mathbf{m}}\:|n\rangle_{\textrm{e}}\otimes
|\mathbf{m}\rangle_\text{ph} \:,
\end{equation}
where the information about the phonon content of this state is contained in the coefficients $C_{n,\mathbf{m}}$.

The truncation of the phonon Hilbert space follows a well-established procedure in which the system size ($N$)
and maximum number of phonons retained ($M$) are gradually increased until the convergence for the ground-state energy 
and phonon distribution is reached~\cite{Wellein+Fehske}. The convergence criterion adopted here is that the 
relative error in these quantities upon further increase of $N$ and $M$ is not larger than $10^{-4}$. While 
for Holstein-type coupling the system size is practically inconsequential~\cite{Ranninger:92}, this is not the 
case for the nonlocal Peierls-type coupling investigated here. In particular, the stated criterion is here 
satisfied for a system with $N=6$ sites and $M=8$ phonons, the values adopted in the following.
\section{Entanglement spectrum} \label{entspectrum}
To set the stage for further discussion, the concept of entanglement spectra for bipartite quantum systems 
is briefly introduced here, complemented by its specific application to the coupled e-ph system under 
consideration. To begin with, a reminder is presented about some basic aspects of entanglement in 
bipartite systems, including the definition of von Neumann entanglement entropy (Sec.~\ref{Bipartite}). 
The most general features of entanglement spectra, exemplified by their intimate connection to the Schmidt 
decomposition~\cite{Schmidt:1907,Ekert+Knight:95}, are then briefly reviewed (Sec.~\ref{EntSpectrum}). 
Finally, Sec.~\ref{symmconsid} is devoted to general considerations on labeling the entanglement-spectrum 
eigenvalues with quantum numbers of certain symmetry-related observables, as well as their concrete 
use in the coupled e-ph system at hand. 
\subsection{Bipartite systems, entanglement entropy, and application to the coupled e-ph system} \label{Bipartite}
The Hilbert space of a quantum system that can be divided up into two subsystems $A$ and $B$ has the form of a tensor product 
$\mathcal{H}=\mathcal{H}_{\textrm{A}}\otimes\mathcal{H}_{\textrm{B}}$ of the component spaces. In what follows the respective 
dimensions of $\mathcal{H}_{\textrm{A}}$ and $\mathcal{H}_{\textrm{B}}$ will be denoted by $d_{\textrm{A}}$ and $d_{\textrm{B}}$.

In a pure state $|\Psi\rangle$ -- not necessarily normalized -- the density matrix of the full system is given by
\begin{equation}\label{rhogen}
\rho=\frac{|\Psi\rangle\langle\Psi|}{\langle\Psi|\Psi\rangle}  \:.
\end{equation}
The reduced (marginal) density matrix $\rho_{\textrm{\tiny{A}}}$ of the subsystem 
$A$ is obtained by tracing $\rho$ over the degrees of freedom of the subsystem $B$: 
$\rho_{\textrm{A}}=\tr_{\textrm{B}}\rho$.
The von Neumann (entanglement) entropy, defined as
\begin{equation} \label{vonNeumannS}
S_{\textrm{E}}= -\textrm{Tr}_{\textrm{A}}(\rho_{\textrm{A}}\ln\rho_{\textrm{A}}) \:,
\end{equation}
describes the quantum correlations in the state $|\Psi\rangle$. Note that 
$S_{\textrm{E}}=-\tr_{\textrm{A}}(\rho_{\textrm{A}}\ln\rho_{\textrm{\tiny{A}}})=
-\tr_{\textrm{B}}(\rho_{\textrm{\tiny{B}}}\ln\rho_{\textrm{B}})$, where the reduced density 
matrix $\rho_{\textrm{B}}$ is obtained by tracing $\rho$ over the degrees of freedom of the 
subsystem $A$. 

In accordance with general relation in Eq.~\eqref{rhogen}, the density matrix corresponding to the ground
state $|\psi_{\textrm{gs}}\rangle$ of the coupled e-ph system ($A\rightarrow\textrm{e}$, $B\rightarrow \textrm{ph}$) 
with the tensor-product Hilbert space ${\mathcal H}={\mathcal H}_{\textrm{e}}\otimes{\mathcal H}_{\textrm{ph}}$ is 
given by
\begin{equation}\label{rho_eph}
\rho_{\textrm{e-ph}}=\frac{|\psi_{\textrm{gs}}\rangle
\langle\psi_{\textrm{gs}}|}{\langle\psi_{\textrm{gs}}|\psi_{\textrm{gs}}\rangle} \:.
\end{equation}
The reduced excitation density matrix is then given by
\begin{equation} \label{rho_e}
\rho_{\textrm{e}}= \textrm{Tr}_{\textrm{ph}}\big(\rho_{\textrm{e-ph}}\big) \:,
\end{equation}
and the ground-state entanglement entropy $S_{\textrm{gs}}$ of the system is defined as
\begin{equation} \label{S_gs}
S_{\textrm{gs}}= -\textrm{Tr}_{\textrm{e}}\big(\rho_{\textrm{e}}\ln
\rho_{\textrm{e}}\big) \:.
\end{equation}
\subsection{Entanglement spectrum: generalities}\label{EntSpectrum}
Let $\{|s_{\textrm{A}}\rangle, s_{\textrm{A}}=1,\ldots,d_{\textrm{A}}\}$ and 
$\{|s_{\textrm{B}}\rangle, s_{\textrm{B}}=1,\ldots,d_{\textrm{B}}\}$ be orthonormal bases in the component 
spaces $\mathcal{H}_{\textrm{A}}$ and $\mathcal{H}_{\textrm{B}}$ of the above bipartite system.
A generic pure quantum state $|\Psi\rangle$ of the bipartite system can be decomposed in the orthonormal 
basis $\{|s_{\textrm{A}}\rangle \otimes |s_{\textrm{B}}\rangle\}$, i.e., the tensor product of 
$\{|s_{\textrm{A}}\rangle\}$ and $\{|s_{\textrm{B}}\rangle\}$:
\begin{equation}\label{statePsi}
|\Psi\rangle = \sum_{s_{\textrm{A}}=1}^{d_{\textrm{A}}}\sum_{s_{\textrm{B}}=1}
^{d_{\textrm{B}}}\:c_{s_{\textrm{A}},s_{\textrm{B}}}\:
|s_{\textrm{A}}\rangle\otimes|s_{\textrm{B}}\rangle \:.
\end{equation}
The coefficients $c_{s_{\textrm{A}},s_{\textrm{B}}}$ in this last expansion can be thought of as the matrix 
elements of a (generically rectangular) matrix $M$, which will henceforth be referred to as the entanglement 
matrix. Through singular-value decomposition (SVD) this matrix can be recast as 
\begin{equation}\label{SVDentMatrix}
M = UDV^{\dagger} \:,
\end{equation}
where $U$ is a matrix of dimension $d_{\textrm{A}}\times \textrm{min} (d_{\textrm{A}},d_{\textrm{B}})$ that satisfies 
$U^{\dagger}U = \mathbbm{1}$ and $V$ a $d_{\textrm{B}}\times \textrm{min}(d_{\textrm{A}},d_{\textrm{B}})$ matrix which 
satisfies $VV^{\dagger}=\mathbbm{1}$; $D$ is a diagonal square matrix of dimension $\textrm{min} (d_{\textrm{A}},d_{\textrm{B}})$ 
where all entries -- the singular values of the matrix $M$ -- are non-negative and can be written as $\{e^{-\xi_{\alpha}/2}|\:
\alpha=1,\ldots,\textrm{min} (d_{\textrm{A}},d_{\textrm{B}})\}$.

Using the above SVD of the entanglement matrix, one arrives at the Schmidt decomposition~\cite{Ekert+Knight:95}
\begin{equation}\label{SchmidtDecomp}
|\Psi\rangle = \sum_{\alpha=1}^{\alpha_{\textrm{max}}}\:e^{-\xi_{\alpha}/2}
|\psi^{\alpha}_\textrm{A}\rangle\otimes|\psi^{\alpha}_\textrm{B}\rangle \:,
\end{equation}
where $\alpha_{\textrm{max}}=\textrm{min}(d_{\textrm{A}},d_{\textrm{B}})$ and
\begin{equation} \label{psiAB}
|\psi^{\alpha}_\textrm{A}\rangle = \sum_{s_{\textrm{A}}=1}^{d_{\textrm{A}}}\:U^{\dagger}_{\alpha,s_{\textrm{A}}} 
|s_{\textrm{A}}\rangle \:,\quad
|\psi^{\alpha}_\textrm{B}\rangle = \sum_{s_{\textrm{B}}=1}^{d_{\textrm{B}}}\:V^{\dagger}_{\alpha,s_{\textrm{B}}} 
|s_{\textrm{B}}\rangle \:,
\end{equation}
are the singular vectors of the matrix $M$. The latter allow one to express the reduced density matrices as
\begin{eqnarray} \label{rhoArhoB}
\rho_{\textrm{A}} = \sum_{\alpha=1}^{\alpha_{\textrm{max}}}\:e^{-\xi_{\alpha}}|\psi^{\alpha}_{\textrm{A}} 
\rangle\langle\psi^{\alpha}_{\textrm{A}}| \:, \nonumber \\
\rho_{\textrm{B}} = \sum_{\alpha=1}^{\alpha_{\textrm{max}}}\:e^{-\xi_{\alpha}}|\psi^{\alpha}_{\textrm{B}} 
\rangle\langle\psi^{\alpha}_{\textrm{B}}| \:.
\end{eqnarray}
Thus the joint spectrum of $\rho_{\textrm{A}}$ and $\rho_{\textrm{B}}$ can be obtained from the Schmidt decomposition of 
the state $|\Psi\rangle$ [cf. Eq.~\eqref{SchmidtDecomp}] (or, equivalently, from the SVD of the entanglement matrix) 
and is given by the set $\{e^{-\xi_{\alpha}}\}$ (i.e., squares of the above singular values). In particular, the {\em 
entanglement spectrum} corresponds to the set $\{\xi_{\alpha}\}$ of the negative logarithms of the joint eigenvalues of 
$\rho_{\textrm{A}}$ and $\rho_{\textrm{B}}$. 
\subsection{Symmetry-related considerations and application to the coupled e-ph system}\label{symmconsid}
In what follows, it is shown that the entanglement-spectrum eigenvalues of the e-ph system can be labeled 
by the quantum number associated with the excitation quasimomentum operator, this being a special case of 
more general symmetry-related considerations.

Consider a Hermitian operator (observable) $\mathcal{O}$ acting on the tensor-product Hilbert space $\mathcal{H}=\mathcal{H}_{\textrm{A}}
\otimes\mathcal{H}_{\textrm{B}}$ that can be decomposed as $\mathcal{O}=\mathcal{O}_{\textrm{A}}+\mathcal{O}_{\textrm{B}}$, 
where $\mathcal{O}_{\textrm{A}}$ acts only on $\mathcal{H}_{\textrm{A}}$ and $\mathcal{O}_{\textrm{B}}$ only on 
$\mathcal{H}_{\textrm{B}}$. Assuming that the state $|\Psi\rangle$ is an eigenstate of $\mathcal{O}$, it immediately 
follows that its corresponding density matrix $\rho$ commutes with $\mathcal{O}$. Furthermore, $[\mathcal{O},\rho]=0$ 
implies that $\textrm{Tr}_{\textrm{B}}[\mathcal{O},\rho]=\textrm{Tr}_{\textrm{B}}[\mathcal{O}_{A},\rho]+
\textrm{Tr}_{\textrm{B}}[\mathcal{O}_{\textrm{B}},\rho]=0$. By virtue of the fact that $\textrm{Tr}_{\textrm{B}}
[\mathcal{O}_{\textrm{B}},\rho]=0$, which can be verified by evaluating this last trace in the
eigenbasis of the operator $\mathcal{O}_{\textrm{B}}$, and 
\begin{equation}
\textrm{Tr}_{\textrm{B}}[\mathcal{O}_{\textrm{A}},\rho]=[\mathcal{O}_{\textrm{A}},\textrm{Tr}_{\textrm{B}}
\rho]\equiv [\mathcal{O}_{\textrm{A}},\rho_{\textrm{A}}]\:,
\end{equation}
one readily finds that $[\mathcal{O}_{\textrm{A}},\rho_{\textrm{A}}]=0$. Therefore, one can simultaneously 
diagonalize $\rho_{\textrm{A}}$ and $\mathcal{O}_{\textrm{A}}$, and label the entanglement-spectrum eigenvalues 
$\{\xi_{\alpha}\}$ according to the quantum number of $\mathcal{O}_{\textrm{A}}$.

It is pertinent to apply these general symmetry-related considerations to the coupled e-ph system at hand, 
which possesses a discrete translational symmetry. Owing to this symmetry, mathematically expressed by 
$[H,K_{\textrm{tot}}]=0$, the ground state $|\psi_{\textrm{gs}}\rangle$ of the system is an eigenstate of 
the operator $K_{\textrm{tot}}$ [cf. Eq.~\eqref{totalcryst}]. This operator -- the generator of discrete 
translations -- plays the role of the observable $\mathcal{O}$ in the above discussion. Namely, it can be 
decomposed as $K_{\textrm{tot}}=K_{\textrm{e}}+K_{\textrm{ph}}$, where $K_{\textrm{e}}=\sum_{k} k\:c^{\dagger}_{k}c_{k}$ 
acts only on $\mathcal{H}_{\textrm{e}}$ and $K_{\textrm{ph}}=\sum_{q}q\:b^{\dagger}_{q}b_{q}$ on 
$\mathcal{H}_{\textrm{ph}}$.  Following the above general reasoning, one concludes that the operator 
$K_{\textrm{e}}$ commutes with the reduced density matrix $\rho_{\textrm{e}}$ 
corresponding to $|\psi_{\textrm{gs}}\rangle$ [cf. Eq.~\eqref{rho_e}]. Thus, the operators $K_{\textrm{e}}$ 
and $\rho_{\textrm{e}}$ can be diagonalized simultaneously and the entanglement-spectrum eigenvalues $\{\xi_1,\ldots,\xi_N\}$ 
can be labeled by the quantum number of $K_{\textrm{e}}$, i.e., they correspond to different quasimomenta 
in the Brillouin zone permissible by the periodic boundary conditions (cf. Sec.~\ref{ModelHamiltonian}). In 
particular, the excitation-quasimomentum eigenvalue $K^{\alpha}_{\textrm{e}}\equiv\langle\xi_{\alpha}|\:K_{\textrm{e}}\:|\xi_{\alpha}\rangle$ 
corresponding to $\xi_{\alpha}$ ($\alpha=1,\ldots,N$) is given by Eq.~\eqref{FinalKe} in Appendix~\ref{KeExpr}.
\section{Results and Discussion} \label{ResultsDiscuss}
The strategy employed here to analyze the coupled e-ph system entails the following steps. After 
the ground-state vector $|\psi_{\textrm{gs}}\rangle$ -- represented by the coefficients $C^{\textrm{gs}}_{n,\mathbf{m}}$ 
[cf. Eq.~\eqref{expandPsi}] -- is obtained through Lanczos diagonalization~\cite{CullumWilloughbyBook,PrelovsekBoncaChapter:13} 
of the e-ph Hamiltonian \eqref{Hamiltonian} for each value of $\lambda_{\textrm{eff}}$ in the chosen range $[0,4]$, the reduced 
density matrix is determined with the aid of Eqs.~\eqref{rho_eph} and \eqref{rho_e}. Its matrix elements 
$(\rho_{\textrm{e}})_{nn'}$ ($n,n'=1,\ldots,N$) are given by
\begin{equation}\label{rho_e}
(\rho_{\textrm{e}})_{nn'} = \frac{\displaystyle\sum_{\mathbf{m}}\:C^{\textrm{gs}}_{n,\mathbf{m}}C^{\textrm{gs}\:*}
_{n',\mathbf{m}}}{\displaystyle\sum^{N}_{p=1}\sum_{\mathbf{m}}\:|C^{\textrm{gs}}_{p,\mathbf{m}}|^{2}} \:.
\end{equation}
The entanglement-spectrum eigenvalues and their associated eigenvectors are then obtained by simply solving the 
($N\times N$)-dimensional eigenproblem of $\rho_{\textrm{e}}$. Alternatively, the same spectrum can be obtained
through a numerical SVD~\cite{NRcBook} of the corresponding entanglement matrix [cf. Eq.~\eqref{SVDentMatrix}]. 
\begin{figure}[b!]
\includegraphics[clip,width=8.45cm]{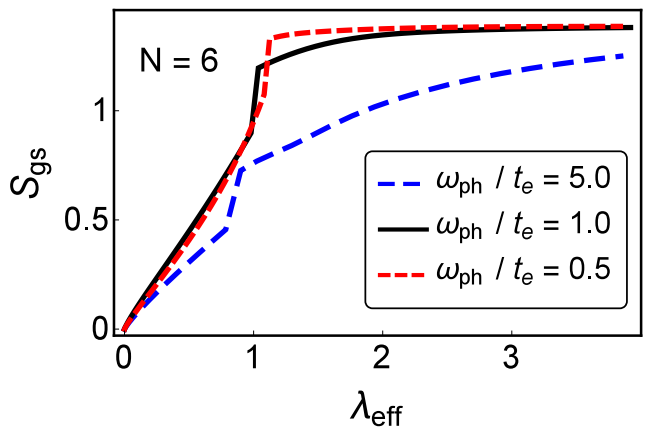}
\caption{\label{fig:EntropyPlot}Dependence of the ground-state e-ph entanglement entropy for a system of size 
$N=6$ on the effective coupling strength, depicted for three different values of the adiabaticity ratio.}
\end{figure}

In what follows, the entire range of e-ph coupling strengths is analyzed -- from the weak-coupling regime characterized by a 
weakly-dressed (quasi-free) excitation to the strong-coupling regime with a heavily-dressed one (small polaron). The analysis 
was repeated for different values of the adiabaticity ratio, covering the adiabatic ($\omega_{\textrm{ph}}/t_{\rm e}<1$) and 
antiadiabatic ($\omega_{\textrm{ph}}/t_{\rm e}>1$) regimes, as well as the intermediate case ($\omega_{\textrm{ph}}/t_{\rm e}=1$). 
\begin{figure}[t!]
\includegraphics[clip,width=8.45cm]{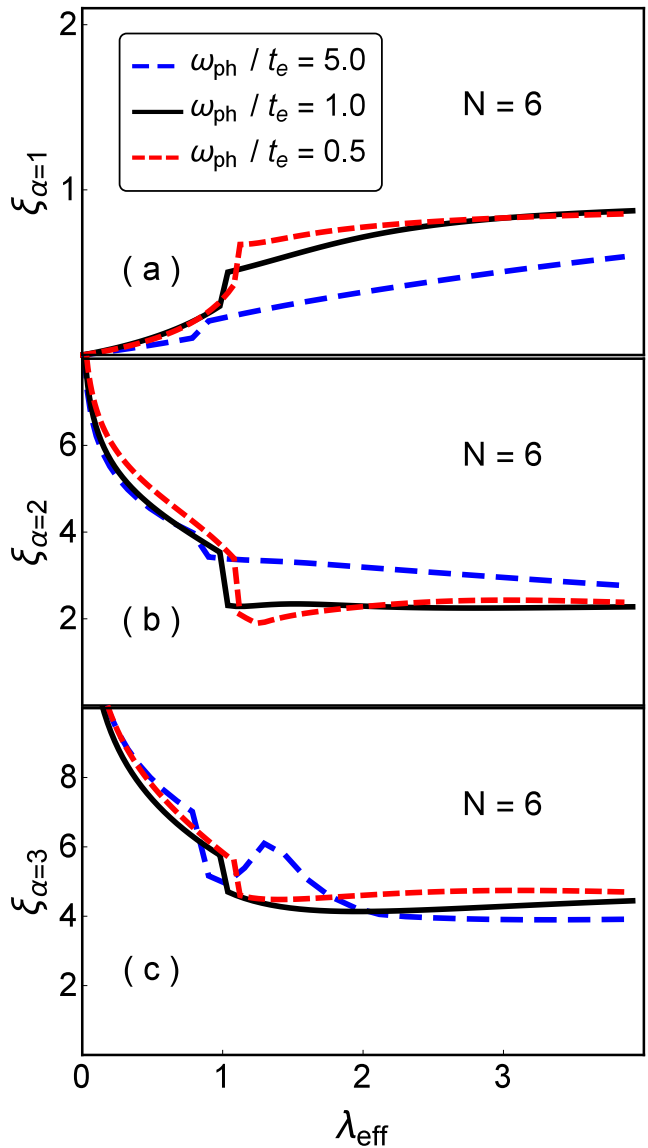}
\caption{\label{fig:xi123Plot}Entanglement-spectrum eigenvalue $\xi_{\alpha}$ in the ground state of a system of size $N=6$ 
as a function of the effective coupling strength: (a) $\alpha=1$, (b) $\alpha=2$, and (c) $\alpha=3$.}
\end{figure}

Before embarking on the analysis of the ground-state entanglement spectrum of the system it is instructive to discuss its corresponding
entanglement entropy $S_{\textrm{gs}}$ [cf. Eq.~\eqref{S_gs}]. In Fig.~\ref{fig:EntropyPlot}, this quantity is depicted for three different 
values of the adiabaticity ratio and clearly shows a first-order nonanalyticity at a critical value $\lambda^{\textrm{c}}_{\textrm{eff}}$ 
of $\lambda_{\textrm{eff}}$. This critical value decreases -- albeit rather slowly -- with $\omega_{\textrm{ph}}/t_{\rm e}$. Beyond this 
critical value, the entanglement entropy grows monotonously and for a sufficiently large coupling strength saturates at the value $\ln N$ 
characteristic of maximally-entangled states~\cite{Zhao+:04} in this system; for $N=6$, this maximal value is $S^{\textrm{max}}_{\textrm{gs}}
\approx 1.79$ [cf. Fig.~\ref{fig:EntropyPlot}].

\begin{figure}[b!]
\includegraphics[clip,width=8.45cm]{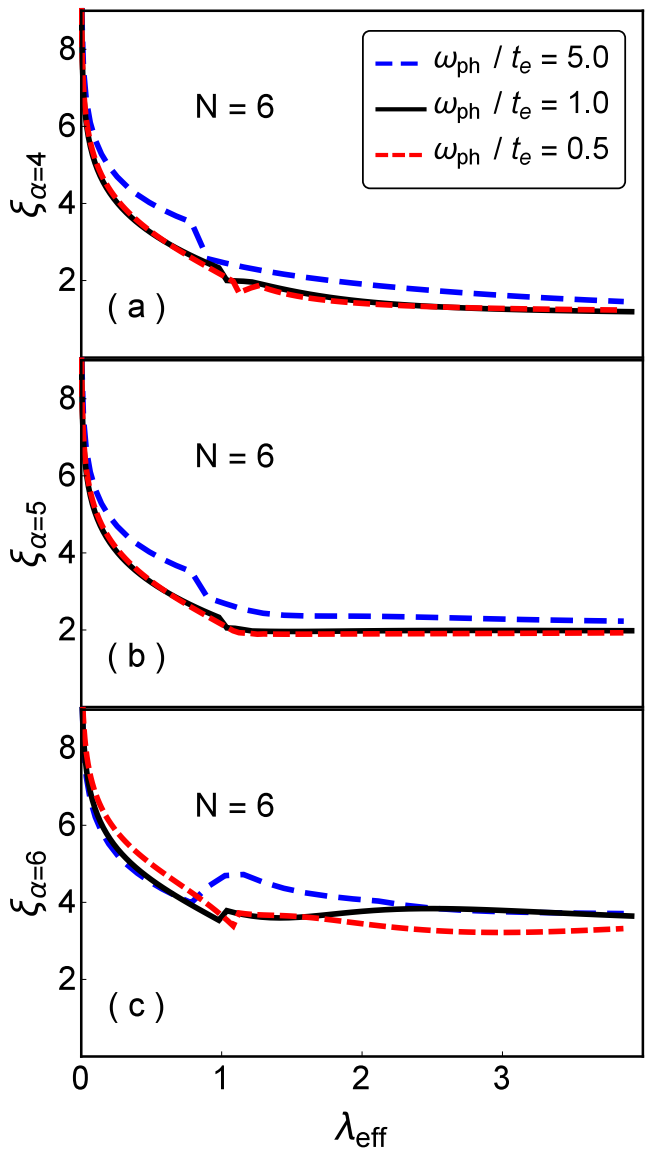}
\caption{\label{fig:xi456Plot}Entanglement-spectrum eigenvalue $\xi_{\alpha}$ in the ground state of a system of size $N=6$ 
as a function of the effective coupling strength: (a) $\alpha=4$, (b) $\alpha=5$, and (c) $\alpha=6$.}
\end{figure}

The numerically-obtained entanglement spectrum has the same qualitative structure for all values of the adiabaticity ratios, which 
appears to be consistent with the previously established general conclusion that the gross features of small polarons in the presence 
of Peierls-type coupling are for the most part insensitive to the value of $\omega_{\textrm{ph}}/t_{\rm e}$~\cite{Capone+:97}. Its 
corresponding eigenvalues, i.e., their dependence on $\lambda_{\textrm{eff}}$, are depicted in Figs.~\ref{fig:xi123Plot} ($\alpha=1,2,3$) 
and \ref{fig:xi456Plot} ($\alpha=4,5,6$) for all three relevant regimes. While the nonanalytic behavior is manifested in all six eigenvalues,
what is noticeable from Figs.~\ref{fig:xi123Plot} and \ref{fig:xi456Plot} is that this nonanalyticity is much more pronounced in the three 
eigenvalues shown in Fig.~\ref{fig:xi123Plot} than in those displayed in Fig.~\ref{fig:xi456Plot}.

\begin{figure}[t!]
\includegraphics[clip,width=8.45cm]{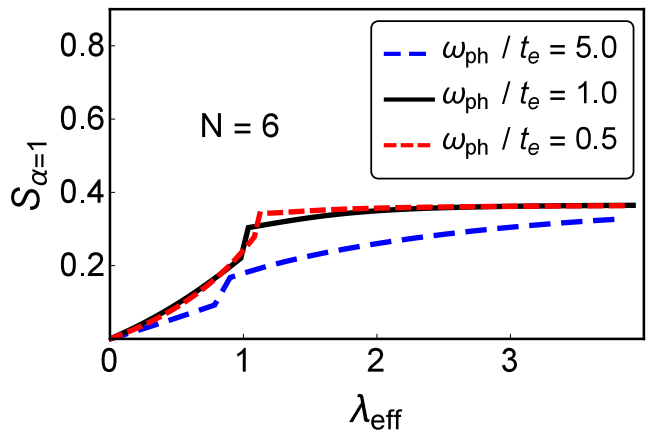}
\caption{\label{fig:Salpha1Plot}Contribution $S_{\alpha=1}\equiv \xi_{\alpha=1}\:e^{-\xi_{\alpha=1}}$ of the $\alpha=1$
entanglement-spectrum eigenvalue to the ground-state entanglement entropy $S_{\textrm{gs}}$.}
\end{figure}

Importantly, from Fig.~\ref{fig:xi123Plot} it can be inferred that the behavior of the ground-state entanglement entropy 
$S_{\textrm{gs}}=\sum^{6}_{\alpha=1}\xi_{\alpha}\:e^{-\xi_{\alpha}}$ [displayed in Fig.~\ref{fig:EntropyPlot}] -- especially 
for $\lambda_{\textrm{eff}}\ge\lambda^{\textrm{c}}_{\textrm{eff}}$ -- is determined almost entirely by that of the smallest 
entanglement-spectrum eigenvalue ($\alpha=1$)[cf. Fig.~\ref{fig:xi123Plot}(a)], i.e., the largest eigenvalue of the corresponding 
reduced density matrix [cf. Eq.~\eqref{rho_e}]. Namely, the remaining five eigenvalues -- especially those corresponding 
to $\alpha=2$ and $\alpha=4$, depicted in Figs.~\ref{fig:xi123Plot}(b) and ~\ref{fig:xi456Plot}(b), respectively -- have 
a rather weak dependence on $\lambda_{\textrm{eff}}$ beyond the critical coupling strength, thus giving nearly constant 
contributions to $S_{\textrm{gs}}$ for $\lambda_{\textrm{eff}}\ge\lambda^{\textrm{c}}_{\textrm{eff}}$. Another  feature that 
sets the $\alpha=1$ eigenvalue apart is that it is the only one which monotonously increases with $\lambda_{\textrm{eff}}$ 
below $\lambda^{\textrm{c}}_{\textrm{eff}}$, with all the other eigenvalues showing fairly similar decreasing behavior for 
$\lambda_{\textrm{eff}}<\lambda^{\textrm{c}}_{\textrm{eff}}$. Interestingly, not only that the $\lambda_{\textrm{eff}}$-dependence of its 
corresponding contribution $S_{\alpha=1}\equiv\xi_{\alpha=1}\:e^{-\xi_{\alpha=1}}$ (cf. Fig.~\ref{fig:Salpha1Plot}) mimics the behavior 
of the total ground-state entanglement entropy $S_{\textrm{gs}}$, but this entanglement-spectrum eigenvalue itself also shows a very similar 
dependence on $\lambda_{\textrm{eff}}$ as $S_{\alpha=1}$ and $S_{\textrm{gs}}$.

This last finding that the ground-state e-ph entanglement entropy $S_{\textrm{gs}}$ is to a large extent determined by 
$\xi_{\alpha=1}$ -- i.e., by the smallest eigenvalue of the corresponding entanglement Hamiltonian -- squares with a conclusion 
drawn in studies of other types of many-body systems. Namely, the interesting, universal part of the entanglement spectrum is typically 
captured by the largest eigenvalues of the relevant reduced density matrix~\cite{Johri+:17}. Recalling that the entanglement 
entropy corresponding to a certain reduced density matrix is equal to the thermodynamic entropy of the attendant entanglement 
Hamiltonian $H_{\textrm{E}}$ at the inverse temperature $\beta_{\textrm{E}}=1$, this finding also becomes closely related to 
the far more general issue as to when an entire Hamiltonian of a many-body system can be considered as being encoded in a single
eigenstate (typically its ground state). Such situations are not uncommon in many-body systems, but have so far been systematically 
discussed only in the context of thermodynamic and entanglement entropies of single-component systems, such as interacting quantum 
spin-$1/2$ chains or interacting hard-core bosons on a 1D lattice~\cite{Garrison+Grover:18}. The present study of the 
entanglement spectrum in a (two-component) coupled e-ph system thus provides another, qualitatively different, example of a physical 
system where this same issue becomes relevant.

\begin{figure}[t!]
\includegraphics[clip,width=8.45cm]{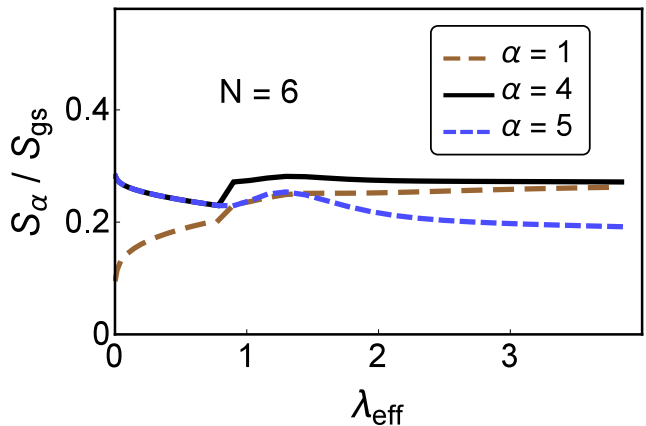}
\caption{\label{fig:Salpha1Sgs}Relative contributions $S_{\alpha}/S_{\textrm{gs}}$ of the entanglement-spectrum 
eigenvalues $\alpha=1,4,5$ to the total ground-state entanglement entropy (independent of the adiabaticity ratio).}
\end{figure}

As regards the relative importance of different entanglement-spectrum eigenvalues, a useful insight can be gleaned by evaluating the relative
contributions $S_{\alpha}/S_{\textrm{gs}}$ of those eigenvalues to the total entanglement entropy at different coupling strengths. The actual 
calculation shows that the eigenvalues $\alpha=1, 4$, and $5$ give much larger contributions to $S_{\textrm{gs}}$ than the remaining ones. To
be more specific, they account for around $80\%$ of $S_{\textrm{gs}}$, with their maximal contributions being attained in the vicinity of the 
critical coupling strength. Their individual relative contributions, depicted in Fig.~\ref{fig:Salpha1Sgs}, are completely 
independent of the adiabaticity ratio (hence the value of $\omega_{\textrm{ph}}/t_{\rm e}$ is not indicated in the plot).

As discussed in Sec.~\ref{symmconsid}, resulting from the presence of a discrete translational symmetry is the possibility to label 
the entanglement-spectrum eigenvalues by the quantum number of the excitation-quasimomentum operator $K_{\textrm{e}}$; its values are 
the quasimomenta $k_n$ in the Brillouin zone permitted by the periodic boundary conditions. Based on the expression given by Eq.~\eqref{FinalKe} 
in Appendix~\ref{KeExpr}, it is straightforward to numerically determine the quasimomenta associated to different eigenvalues $\xi_{\alpha}$
for different coupling strengths and adiabaticity ratios. 

\begin{figure}[t!]
\includegraphics[clip,width=8.45cm]{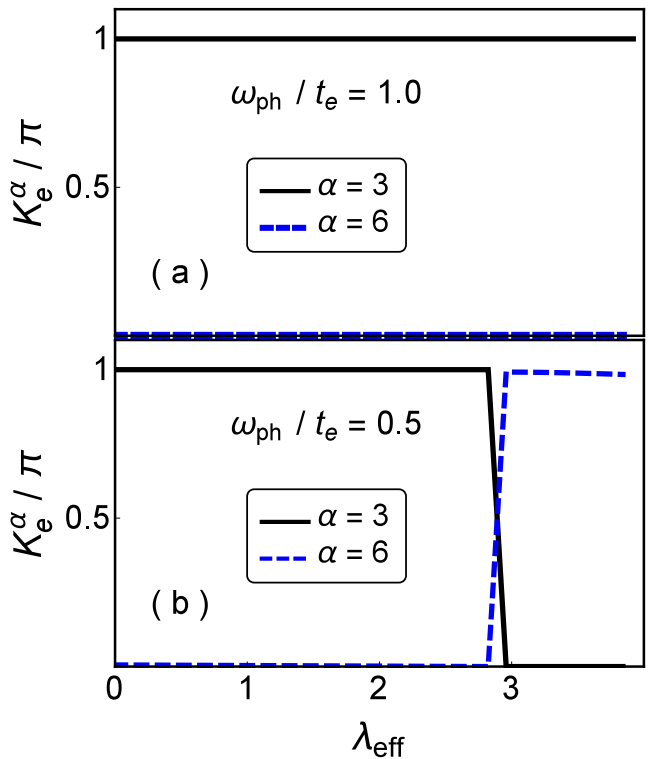}
\caption{\label{fig:Kalpha}Quasimomentum $K^{\alpha}_{\textrm{e}}\equiv \langle\xi_{\alpha}|\:K_{\textrm{e}}\:|\xi_{\alpha}\rangle$ 
(expressed in units of $\pi$) associated to the $\alpha=3$ and $\alpha=6$ entanglement-spectrum eigenvalues 
for (a) $\omega_{\textrm{ph}}/t_{\rm e}=1.0$, and (b) $\omega_{\textrm{ph}}/t_{\rm e}=0.5$.}
\end{figure}

The actual calculation shows that for $\omega_{\textrm{ph}}/t_{\rm e}\geq 1$ (i.e., in the antiadiabatic and intermediate cases) one eigenvalue, 
more precisely $\alpha=3$, corresponds to the quasimomentum $\pi$ at all coupling strenghts, while the five remaining eigenvalues correspond 
to $0$. This is illustrated in Fig.~\ref{fig:Kalpha}(a) for the special case $\omega_{\textrm{ph}}/t_{\rm e}=1$. The corresponding behavior 
for $\omega_{\textrm{ph}}/t_{\rm e}<1$, i.e., in the adiabatic regime, has an additional interesting feature. Namely, while in this regime 
there are eigenvalues corresponding to the bare-excitation quasimomenta $0$ and $\pi$ at all coupling strengths, one also finds cases where a 
specific eigenvalue corresponds to $0$ in a certain interval of coupling strengths and to $\pi$ otherwise. For instance, Fig.~\ref{fig:Kalpha}(b)
illustrates one such example for $\omega_{\textrm{ph}}/t_{\rm e}=0.5$, where for a certain coupling strength slightly below 
$\lambda_{\textrm{eff}}=3$ -- thus lying deeply in the strong e-ph coupling regime -- such a transition occurs between the quasimomenta $0$ 
and $\pi$ for the $\alpha=3$ and $\alpha=6$ entanglement-spectrum eigenvalues.

The occurrence of this last generalized transition provides a differentiation between the e-ph entanglement pattern in the adiabatic regime 
and the other relevant regimes (antiadiabatic, intermediate). This can be linked to the fact that this transition takes 
place at a coupling strength for which maximally-entangled small-polaron states are still not reached in the adiabatic case, unlike in the 
other two cases [cf. Fig.~\ref{fig:EntropyPlot}]. An immediate question is whether a concrete physical meaning can be attributed to it, this 
being related to the much more general issue as to how universal is the entanglement spectrum~\cite{Chandran+:14}. In Ref.~\onlinecite{Chandran+:14}, 
based on several physical examples it was argued that the entanglement Hamiltonian of a physical system may undergo transitions in which 
its ground state and low-energy spectrum exhibit singular changes, even when the 
system actually remains in the same phase. In other words, the entanglement spectrum may exhibit spurious quantum phase transitions that do 
not have any genuine physical counterpart, a property that it shares with the less general concept of entanglement entropy~\cite{Amico:08}. 
While this issue was previously discussed in connection with broken-symmetry or topological phases of many-body systems, here it 
comes up in the qualitatively different context of small-polaron states that do not spontaneously break the discrete translational symmetry 
of the underlying excitation-phonon Hamiltonian.
\section{Summary and Conclusions} \label{SumConcl}
To summarize, in this paper the onset of nonanalytic behavior of ground-state-related properties in models with strongly momentum-dependent 
excitation-phonon coupling was investigated from the point of view of the underlying entanglement spectrum. This was accomplished through 
a case study of a lattice model with Peierls-type coupling whose entanglement spectrum was obtained in a numerically-exact fashion. The 
accompanying analysis was carried out in the full range of the relevant effective excitation-phonon coupling strength -- from weak- (quasifree 
excitation) to strong coupling (heavily-dressed excitation, i.e., small polaron) -- and in different regimes of the adiabaticity ratio.

The main finding of the present work is that the dependence of the ground-state entanglement entropy on the excitation-phonon coupling 
strength -- and, in particular, the first-order nonanalyticity that it shows at the critical coupling strength -- chiefly originates 
from the smallest entanglement-spectrum eigenvalue. Another nontrivial conclusion drawn is that this particular eigenvalue shows a very similar 
dependence on the effective coupling strength as the entanglement entropy itself. In addition, as a special case of quite general symmetry-related 
arguments it was demonstrated that the discrete translational symmetry of the system implies that the entanglement-spectrum eigenvalues can 
be labeled by the bare-excitation quasimomentum quantum number. Finally, it was shown numerically that these eigenvalues are predominantly 
associated to quasimomenta $0$ and $\pi$. Interestingly, it was also found that in particular in the adiabatic regime a generalized transition 
between these two quasimomenta -- for specific entanglement-spectrum eigenvalues -- takes place deeply in the strong-coupling regime. This
feature sets apart the adiabatic regime from the other two relevant regimes.

The present work extends the range of applications of the concept of entanglement spectrum to polaronic systems. Generally speaking, 
what makes the ground-state nonanalyticities in models of the kind investigated here particularly appealing is that they take place in a 
system of finite size and are thus amenable to a rigorous numerical analysis. It would be interesting to test the generality of the conclusions 
drawn here in a future work by studying other models with strongly momentum-dependent excitation-phonon coupling whose ground states show a 
similar nonanalytic behavior. Furthermore, the local (single-qubit) addressability of the previously proposed analog quantum simulators of those 
models~\cite{Stojanovic+:14,Stojanovic+Salom:19} may allow an experimental measurement of the corresponding entanglement spectra. Namely,
a completely general method for such measurements was recently suggested and applied to a specific class of locally-addressable systems (cold 
atoms in optical lattices)~\cite{Pichler+:16}. This method -- based on an analogy to a many-body Ramsey interferometry~\cite{Ekert+:02} -- makes 
use of the fact that the conditional evolution of a many-body system is determined by a copy of its density operator, which acts as the 
Hamiltonian. It is conceivable that the ever-improving scalability and coherence properties of superconducting-qubit systems will allow 
the realization of the aforementioned simulators in not-too-distant future, which will in turn make it possible to measure the relevant
entanglement spectra using the latter method.
\begin{acknowledgments}
The author acknowledges useful discussions on the numerical implementation with I. Salom 
and thanks J. Sous for pointing out Ref.~[46]. This research was supported by the Deutsche 
Forschungsgemeinschaft (DFG) as part of the project S4 within CRC 1119 CROSSING.
\end{acknowledgments}
\appendix 
\section{Derivation of the expression for $\langle\xi_{\alpha}|\:K_{\textrm{e}}\:|\xi_{\alpha}\rangle$} \label{KeExpr}
To begin with, it is worthwhile noting that the expectation value of the operator $K_{\textrm{e}}$ 
with respect to the entanglement-spectrum eigenvector (i.e., entanglement-Hamiltonian eigenstate)
$|\xi_{\alpha}\rangle$ ($\alpha=1,\ldots,N$)
\begin{equation}
\langle \xi_{\alpha}|\:K_{\textrm{e}}\:|\xi_{\alpha}\rangle=\sum_{k}\:k 
\:\langle\xi_{\alpha}|\:c^{\dagger}_{k}c_{k}\:|\xi_{\alpha}\rangle
\end{equation}
can be rewritten as 
\begin{equation} \label{EqExpectKe}
\langle \xi_{\alpha}|\:K_{\textrm{e}}\:|\xi_{\alpha}\rangle=\sum_{k}\:k 
\:\Vert c_{k}\:|\xi_{\alpha}\rangle\Vert^{2}\:,
\end{equation}
where $\Vert \ldots \Vert$ stands for the norm of a vector. On the other hand, the eigenvector 
$|\xi_{\alpha}\rangle$ can be expanded in the basis of the $N$-dimensional excitation Hilbert 
space $\mathcal{H}_{\textrm{e}}$
\begin{equation}
|\xi_{\alpha}\rangle=\sum_{n=1}^{N}\:\xi_{\alpha,n}\:|n\rangle_{\textrm{e}}\equiv 
\sum_{n=1}^{N}\:\xi_{\alpha,n}\:c^{\dagger}_n|0\rangle_{\textrm{e}}\:,
\end{equation}
where $\xi_{\alpha,n}\equiv {}_{\textrm{e}}\langle n|\xi_{\alpha}\rangle$ is the projection of 
$|\xi_{\alpha}\rangle$ onto $|n\rangle_{\textrm{e}}$. By Fourier transforming the momentum-space 
operator $c_{k}$ in Eq.~\eqref{EqExpectKe} back to real space and noting that $c_{n'}c^{\dagger}_n
|0\rangle_{\textrm{e}}\equiv \delta_{n,n'}|0\rangle_{\textrm{e}}$, one readily obtains 
\begin{equation}
c_{k}\:|\xi_{\alpha} \rangle = \frac{1}{\sqrt{N}}\:\sum_{l=1}^{N}
e^{ikl}\:\xi_{\alpha,l}|0\rangle_{\textrm{e}}\:.
\end{equation}
It immediately follows that 
\begin{equation}
\Vert c_{k}\:|\xi_{\alpha}\rangle\Vert^{2}=\frac{1}{N}\:\sum_{l,l'=1}^{N}
e^{ik(l-l')}\:\xi_{\alpha,l}\:\xi^{*}_{\alpha,l'}
\end{equation}
and, finally, by reinserting this last result into Eq.~\eqref{EqExpectKe}, 
\begin{equation}\label{FinalKe}
\langle \xi_{\alpha}|\:K_{\textrm{e}}\:|\xi_{\alpha}\rangle=\frac{1}{N}\:\sum_{k}
\:k\left[\sum_{l,l'=1}^{N}e^{ik(l-l')}\:\xi_{\alpha,l}\:\xi^{*}_{\alpha,l'}\right] \:.
\end{equation}
From this last expression the quasimomenta corresponding to different entanglement-spectrum eigenvalues 
$\xi_{\alpha}$ can easily be determined numerically, using the previously obtained components $\xi_{\alpha,l}$ 
of their corresponding eigenvectors $|\xi_{\alpha}\rangle$ and carrying out the $k$ summation over the
$N$ permissible quasimomenta in the Brillouin zone (cf. Sec.~\ref{ModelHamiltonian}).

\end{document}